\documentclass[letterpaper,12pt]{article}

\usepackage{fullpage}
\usepackage{amssymb}
\usepackage{hyperref}
\usepackage{graphicx}

\begin{document}

\begin{center}
\LARGE\bfseries
Experimental limits on the fundamental Planck scale in large extra
dimensions
\end{center}  

\bigskip

\begin{center}
Douglas M. Gingrich\\

\bigskip

\textit{Centre for Particle Physics, Department of Physics, University
of Alberta, \\ Edmonton, AB T6G 2E1 Canada}\\  
\textit{TRIUMF, Vancouver, BC V6T 2A3 Canada}\\ 
{\footnotesize gingrich@ualberta.ca}
\end{center}

\begin{center}
\small \today
\end{center}

\begin{quotation} \noindent
\textbf{Abstract\ } 
I present an up to date set of limits on the fundamental Planck scale
$M_D$. 
The best limit for each number of extra dimensions $n$ is shown in
bold font.
For 
$n=2$, $M_D > 5.6$~TeV; 
$n=3$, $M_D > 4.4$~TeV; 
$n=4$, $M_D > 3.9$~TeV;
$n=5$, $M_D > 3.6$~TeV;
$n=6$, $M_D > 3.3$~TeV;
and for $7\le n \le 8$, $M_D \gtrsim 0.8$~TeV.  
\end{quotation}

\begin{quotation} \noindent
\end{quotation}

\section{Introduction\label{sec1}}

For black hole studies, we are interested in the $D$-dimensional
fundamental Planck scale $M_D$. 
This scale is related to the derived Planck scale $M_\mathrm{Pl}$ in
models of large extra dimensions by

\begin{equation}\label{eq1}
M_\mathrm{Pl}^2 = 8 \pi R^n M_D^{2+n}\, ,
\end{equation}

\noindent
where $n = D - 4$ is the number of extra dimensions of the same size $R$.
The four-dimensional effective Planck scale is given by Newton's
constant $G_\mathrm{N} = M_\mathrm{Pl}^{-2}$.

Using $M_\mathrm{Pl} = 1.22090\times 10^{16}$~TeV and
$\hbar c = 1.97326968\times 10^{-4}$~TeV-fm gives

\begin{equation}
M_D = \left[ 5.93089 \left( \frac{197.326968~\mu\mathrm{m}}{R} \right)^n
\right]^\frac{1}{n+2} \times 10^\frac{15(2-n)}{n+2}~\mathrm{TeV} 
\end{equation}

\noindent
and

\begin{equation}
R = \left[ \frac{5.93089~\mathrm{TeV}^{n+2}}{M_D^{n+2}}
  \right]^\frac{1}{n} \times 197.326968 \times
  10^\frac{15(2-n)}{n}~\mu\mathrm{m}\, .  
\end{equation}

Limits on $M_D$ or $R$ have been set by direct gravity measurements,
experiments at accelerators, and constraints from astrophysics and
cosmology.
The astrophysical and cosmological limits are high, particularly for
two or three extra dimensions.
However, they are based on a number of assumptions so the results are
only order of magnitude estimates. 
Thus, I will not consider further astrophysical or cosmological limits.

\section{Direct gravity measurements\label{sec2}}

The most straightforward observable effect of the large extra dimensions
is the modification of Newton's gravitational attraction law at very
short distances.
Gravity measurements are sensitive to the largest extra dimension.
The E\"{o}t-Washington group constrain the size of the largest extra
dimension to $R \le 44~\mu$m at the 95\% confidence
level~\cite{Kapner:2006si}. 
This completely rules out TeV-scale gravity with one large extra
dimension. 
For two large extra dimensions, they obtain $M_* \ge 3.2$~TeV.
The PDG transforms this into the limit $R < 30~\mu$m, which corresponds
to $M_D > 4.0$~TeV in the case of $n=2$.
The sensitivity to three extra dimensions of equal size is only $M_D >
4\times 10^{-3}$~TeV. 

\section{Limits from accelerator experiments}\label{sec3}

The HERA experiments have set limits on the Kaluza-Klein
ultraviolet-cutoff scale but not on $M_D$.
I thus consider only the results from the LEP, Tevatron, and LHC
collider experiments.  

In e$^+$e$^-$ processes with real graviton emission, the cross section
is directly sensitive to the number of extra dimensions and the
fundamental scale of gravity.
Virtual graviton exchange is sensitive to the ratio $\lambda/M_H$.
$M_H$ is an ultraviolet-cutoff scale, which is not equivalent to
$M_D$ -- but should be of the same order of magnitude -- and $\lambda$
is a coupling constant, which depends on the underlying theory of
gravity.  
In pp collisions, direct graviton emission also depends on $M_D$, while
virtual graviton exchange does not depend on $M_D$. 
The dependence on the ultraviolet-cutoff is more complicated but the
ideas are similar.

\subsection{Combined LEP results}

I consider only the direct graviton emission searches from
LEP2~\cite{ALEPH,Abdallah:2003np,Achard:2003tx}.   
The results from ALEPH, DELPHI, and L3 have been
combined~\cite{LEP}\footnote{Ref.~\cite{LEP} is found at
http://lepexotica.web.cern.ch/LEPEXOTICA/notes/2004-03/ed\_note\_final.ps.gz}
and are shown in the Table~\ref{tab1}.   

\subsection{Tevatron results}

I consider only the direct graviton emission searches from Run II of
the Tevatron (Table~\ref{tab1}).
The latest CDF search is in jets plus missing transverse
energy final states~\cite{Aaltonen:2008hh}.   
They use a $K$-factor (ratio of cross sections as calculated at the
next-to-leading order and leading order) of 1.3.
The latest D{\O} search is in mono-photon and missing transverse energy
final states~\cite{Abazov:2008kp}.
The $K$-factor is include in the uncertainties. 

\subsection{LHC results}

I consider only the direct graviton emission searches from the LHC
experiments (Table~\ref{tab1}). 
Using approximately 20~fb$^{-1}$ of data at 8~TeV pp centre-of-mass
energy, ATLAS~\cite{Aad:2014tda} and CMS~\cite{Khachatryan:2014rwa} have
search results in mono-photon and missing transverse momentum final
states.  
ATLAS~\cite{Aad:2015zva} and CMS~\cite{Khachatryan:2014rra} also have
search results in mono-jet and missing transverse momentum final
states for the same beam energy and luminosity.

\subsection{Black hole searches}

ATLAS and CMS have searched for direct black hole production.
The limits on $M_D$ from the searches are largely model dependent. 
In the case of classical black hole models, the limits on $M_D$ depend
on the threshold production mass $M_\mathrm{th}$, as well as $n$.
In models of quantum black hole production, CMS~\cite{Khachatryan:2015sja}. 
and ATLAS~\cite{Aad:2014aqa} have searched in di-jet events.   
ATLAS has taken $M_\mathrm{th} = M_D$, allowing limits on $M_D$.
ATLAS has also searched for quantum black holes in the $\gamma+$jet,
e$^+$e$^-$, $\mu^+\mu^-$, e+jet, and $\mu+$jet final states.
The limits on $M_\mathrm{th}$ in these channels are less stringent.  
Since the models are speculative, I do not consider them as giving
limits on $M_D$.  

\begin{table}[htb]
\caption{Upper limits on $M_D$ at the 95\% confidence limit.\label{tab1}} 
\begin{center}
\begin{tabular}{|c|c|c|c|c|c|c|c|}\hline
$n$ & \multicolumn{7}{c|}{$M_D$ [TeV]}\\\hline 
& & & & \multicolumn{2}{c|}{Mono-photon} &
\multicolumn{2}{c|}{Mono-jet}\\\cline{5-8}  
& LEP  & CDF  & D{\O} & ATLAS & CMS & ATLAS & CMS\\\hline 
2 & 1.60 & 1.40 & 0.884 & 2.17 &      & 5.25 & \bf 5.61\\
3 & 1.20 & 1.15 & 0.864 & 2.12 & 2.30 & 4.11 & \bf 4.38\\
4 & 0.94 & 1.04 & 0.836 & 2.13 & 2.70 & 3.57 & \bf 3.86\\
5 & 0.77 & 0.98 & 0.820 & 2.14 & 2.20 & 3.27 & \bf 3.55\\
6 & 0.66 & 0.94 & 0.797 & 2.17 & 2.00 & 3.06 & \bf 3.26\\
7 &      &      & \bf 0.797 &  &      &      &         \\
8 &      &      & \bf 0.778 &  &      &      &         \\
\hline
\end{tabular}
\end{center}
\end{table}

\section*{Acknowledgments}

This work was supported in part by the Natural Sciences and Engineering
Research Council of Canada.

\bibliographystyle{atlasBibStyleWithTitle}
\bibliography{limits}

\end{document}